\documentclass[12pt]{article}
\usepackage{hhline}
\usepackage{latexsym,graphicx}
\usepackage{subfigure}
\usepackage{amssymb}
\usepackage{amsmath}
\usepackage{amscd}
\usepackage{amsthm}
\usepackage{float}
\usepackage{xcolor}
\usepackage[left=2cm,top=2.5cm,right=2.5cm,bottom=1.5cm]{geometry}    
\begin{document}
\begin{center}
\large{\bf{A stable flat entropy-corrected FRW universe}} \\
\vspace{10mm}
\normalsize{Nasr Ahmed$^1$ and Sultan Z. Alamri$^2$}\\
\vspace{5mm}
\small{\footnotesize $^1$ Astronomy Department, National Research Institute of Astronomy and Geophysics, Helwan, Cairo, Egypt\footnote{abualansar@gmail.com}} \\
\small{\footnotesize $^2$ Mathematics Department, Faculty of Science, Taibah University, Saudi Arabia\footnote{szamri@taibahu.edu.sa}} \\
\end{center}  
\date{}
\begin{abstract}
In this paper, a general entropy-corrected FRW cosmological model has been presented in which a deceleration-to-acceleration transition occurs according to recent observations. We found that the case for the flat universe ($k=0$), supported by observations, is the most stable one where it successfully passes all stability tests. The stability of the model has been studied through testing the sound speed, the classical and the new nonlinear energy conditions. The model predicts a positive pressure during the early-time decelerating epoch, and a negative pressure during the late-time accelerating epoch in a good agreement with cosmic history and dark energy assumption. We have investigated all possible values of the prefactors $\alpha$ and $\beta$ in the corrected entropy-area relation to find the best values required for a stable flat universe. We have also made use of the evolution of the equation of state parameters $\omega(t)$ in predicting the correct values of $\alpha$ and $\beta$. The jerk and density parameters have been calculated where a good agreement with observations and $\Lambda$CDM model has been obtained. Two dark energy proposals have been investigated in this model, the entropy-corrected holographic dark energy and the modified holographic Ricci dark energy.

\end{abstract}

\section{Introduction and motivation}
A major development in modern cosmology was the discovery of the late-time cosmic acceleration \cite{11,13,14} which also represents a challenge to our understanding of the standard models of gravity. Observations also show that the universe is flat, highly homogeneous and isotropic on large scales \cite{teg,ben,sp}. To explain this accelerating expansion, an exotic form of energy with negative pressure, dubbed as dark energy, has been assumed. Such small positive energy with negative pressure represents a repulsive gravity that can accelerate the expansion. Several approaches have been proposed to explain late-time cosmic acceleration and several dark energy models have been constructed through modified gravity theories \cite{quint}-\cite{ark22} and dynamical scalar fields \cite{1,moddd,39}, \cite{noj1}-\cite{nass22}. \par
Hawking radiation \cite{hawkk} is a quantum phenomenon in which the black hole's entropy $S$ is proportional to its horizon area $A$, the discovery of this phenomenon indicated a deep connection between gravity and thermodynamics \cite{therm}. The horizon entropy $S=\frac{A}{4G}$ and the Hawking temperature $T=\frac{\left|\kappa_{sg}\right|}{2\pi}$ are connected through the first law of black hole thermodynamics $T dS = dE$, where $\kappa_{sg}$ is the surface gravity and $dE$ is the energy change \cite{61,62,63}. In general, we have $dE=T dS+ \it{work~terms}$ where the work terms depend on the type of black hole. This equation indicates the possibility of the thermodynamic interpretation of Einstein equations near horizon, because geometric quantities of black hole solutions to Einstein equations are related to thermodynamic quantities \cite{80,81}. Jacobson \cite{64} derived Einstein equations using Clausius relation $T dS = \delta Q$ and the horizon- entropy area relation, where $\delta Q$ and $T$ are the energy flux across the horizon and Unruh temperature respectively. The so called unified first law of black hole dynamics and relativistic thermodynamics $dE=TdS+WdV$ was derived by Hayward \cite{82} in spherically symmetric space-times, where $W$ is the work density defined by $-\frac{1}{2} T^{ab}h_{ab}$. \par
In cosmology, the FRW cosmological equations have been derived from the first law of thermodynamics by applying Clausius relation to the apparent horizon of the FRW universe \cite{65, bo,boo}. Later \cite{60}, it has been shown that the FRW cosmological equations can be expressed as $dE=TdS+WdV$ at the apparent horizon where $E=\rho V$ is the total energy and $W=\frac{1}{2}(\rho-p)$ is the work density \cite{82}. Here $\rho$ and $p$ denotes the energy density and pressure of cosmic matter, while $T$ and $S$ are temperature and entropy associated with the apparent horizon. The entropy-area formula holds only for General Relativity, and so it needs corrections when some higher order curvature
term appears \cite{basic1}. The question whether it is still possible to derive modified Friedmann equations after the modifications of the entropy-area relation has been discussed in \cite{basic1} where modified Friedmann equations have been derived by applying the corrected entropy-area relation
\begin{equation} \label{ent}
S=\frac{A}{4G}+\alpha \ln \frac{A}{4G}+\beta \frac{4G}{A}.
\end{equation}
Here $\alpha$ and $\beta$ are dimensionless constants whose values are in debate and not yet determined even within loop quantum gravity \cite{salehi}. The correction terms in (\ref{ent}) appear in the black hole entropy in loop quantum gravity due to thermal equilibrium fluctuations, quantum fluctuations, or mass and charge fluctuations (see \cite{echde} and references therein). The second correction term has also appeared in the entropic cosmology model presented in \cite{caii} where a unification of the holographic inflation and late-time acceleration has been suggested. While some works suggest positive or negative values of $\alpha$ and $\beta$ \cite{zer1}-\cite{zer5}, it has been argued in \cite{zer} that the “best guess” might simply be zero. One of the aims of the current work is to find the best possible values for $\alpha$ and $\beta$ required to describe a stable flat universe in which a deceleration-to-acceleration transition occurs according to recent observations. \par
The paper is organized as follows: section 1 is an introduction. In section 2, we present a solution to the modified entropy-corrected cosmological equations and obtain the formulas for the pressure, energy density, EoS parameter and the jerk parameter. We then analyze the evolution of these functions with cosmic time. The stability of the model has been discussed in section 3. In section 4, we calculate the total density parameter using two different proposals for dark energy, and compare the result with observations. The final conclusion is included in section 5.
\section{Cosmological equations and solutions} \label{sol}
Taking into account the corrected entropy-area relation (\ref{ent}), the following modified FRW equations have been derived \cite{basic1}
\begin{eqnarray} 
H^2+\frac{k}{a^2}+\frac{\alpha G}{2 \pi}\left(H^2+\frac{k}{a^2}\right)^2-\frac{\beta G^2}{3\pi^2}\left(H^2+\frac{k}{a^2}\right)^3&=&\frac{8\pi G}{3}\rho. \label{cosm1}\\
2\left(\dot{H}-\frac{k}{a^2}\right)\left(1+\frac{\alpha G}{\pi} \left(H^2+\frac{k}{a^2}\right) - \frac{\beta G^2}{\pi^2}\left(H^2+\frac{k}{a^2}\right)^2  \right)&=&-8\pi G (\rho+p).\label{cosm2}
\end{eqnarray}
Where $k=0,1, -1$ for a flat, closed and open universe respectively. Since recent observations indicate that the universe accelerates after an epoch
of deceleration \cite{11,rrr}, we can explore new solutions through physically reasonable empirical forms of $a(t)$ that allow the deceleration parameter $q$ to change sign from positive (decelerating phase) to negative (accelerating phase):
\begin{equation} \label{ansatz}
a(t)=A\sqrt{\sinh(\xi t)}
\end{equation}
In addition to the agreement with observations, such hyperbolic form of the scale factor appears in many contexts of cosmology. A generalization of the ansatz (\ref{ansatz}), $a(t)=(\sinh(\xi t))^{\frac{1}{n}}$, has been used in the study of Bianchi cosmological models where a good agreement with observations has been obtained \cite{pr}. A Quintessence model with double exponential potential has been constructed in \cite{sen} assuming the form $a(t)= \frac{a_o}{\alpha}[\sinh(t/t_o)]^{\beta}$, where $t_o$ is the present time, $R_o$ is the present day scale factor, $\beta$ is a constant and $\alpha = [\sinh(1)]^{\beta}$. As has been mentioned in \cite{sen}, the main motivation for assuming this form is its consistency with observations as it gives both early-time deceleration and late-time acceleration. For $\beta <1$, the universe is decelerating for $t \ll t_o$ and exponentially accelerating for $t \gg t_o$. For example, setting $\beta=\frac{2}{3}$ gives $a(t) \propto t^{\frac{2}{3}}$ for $t \ll t_o$, and $a(t) \propto e^{t}$ for $t \gg t_o$. In the study of Ricci dark energy in Chern-Simons modified gravity, it has been shown that the evolution of the scale factor is given by $a(t)=\left(\frac{2\zeta}{3c_1}\right)^{\frac{1}{6}} \sinh^{\frac{1}{3}}(3\sqrt{c_1t})$ \cite{sent}, where $\zeta$ and $c_1$ are constants. In the context of $\Lambda CDM$ model, a unified analytic solution is obtained for the scale factor as $a(t)=\left(\frac{\Omega_m}{\Omega_{\Lambda}}\right)^{\frac{1}{3}} \sinh^{\frac{2}{3}}(\frac{3}{2}\sqrt{\Omega_{\Lambda}}H_ot)$, describing the cosmic evolution from the matter-dominated epoch up to the late-time future \cite{sz} where $\Omega_m=0.27$, $\Omega_m=0.73$ and $H_o$ is the current value of the Hubble constant.\par
Since the main of the current work is investigating the true values of the prefactors $\alpha$ and $\beta$ and not the analytical solutions of the system (\ref{cosm1}) and (\ref{cosm2}), exploring the cosmic behaviour through an empirical ansatz represents an alternative way to achieve our task. The use of such scale factor should lead to a stable solution which is a good opportunity to investigate the best allowed values of the prefactors. The same investigation has been done in \cite{nasastro} using a different empirical ansatz where exactly the same results have been reached which represents a great support to the current work. Investigating relation (\ref{ent}) in different cosmological contexts can provide an accurate estimation to the true values of $\alpha$ and $\beta$ on the cosmological scale. Depending on the values of $\alpha$ and $\beta$, The existence of bouncing solutions of the system (\ref{cosm1}) and (\ref{cosm2}) has been discussed in details in \cite{salehi}. The modified Friedmann equations derived from the corrected entropy-area relation without the last $\beta$ term have been derived in (\cite{basic1}) where some possible analytical solutions have been discussed depending on the values of $\alpha$.
Taking (\ref{ansatz}) into account, the expression for deceleration parameter $q$ can now be written as 
\begin{equation} \label{q1}
q(t)=-\frac{\ddot{a}a}{\dot{a}^2}=\frac{-\cosh^2(t)+2}{\cosh^2(t)}
\end{equation}
Solving (\ref{cosm1}) and (\ref{cosm2}) with the ansatz (\ref{ansatz}), the expressions for the pressure $p(t)$, energy density $\rho(t)$, and EoS parameter $\omega(t)$ can be written as 
\begin{eqnarray}
p(t)&=&\frac{1}{512}\frac{1}{\pi^3\sinh^6(t)}[ (-48\pi^2-6\pi \alpha+\beta) \cosh^6(t)  \\ \nonumber 
 &+&( -64(\pi^2+\frac{1}{4}\pi\alpha-\frac{1}{16}\beta)k\sinh(t)+160\pi^2+ (32\alpha k^2+22\alpha)\pi-16\beta k^2-4\beta ) \cosh^4(t) \\ \nonumber 
 &-&(64(-2\pi^2-\frac{5}{4}\pi\alpha+\beta (k^2+\frac{1}{2})) k\sinh(t)-176\pi^2+(-64\alpha k^2-16\alpha)\pi-48\beta k^2) \cosh^2(t) \\ \nonumber 
 &+& 64 k (\beta k^2-\pi^2-\pi \alpha) \sinh(t)+32 \pi \alpha k^2+64 \beta k^2+64 \pi^2]
\end{eqnarray}
\begin{eqnarray}
\rho(t)&=&\frac{1}{512}\frac{1}{\pi^3\sinh^6(t)}[ (48\pi^2+6\pi \alpha-\beta) \cosh^6(t)  \\ \nonumber 
 &+& (192(\pi^2+\frac{1}{4}\pi\alpha-\frac{1}{16}\beta)k\sinh(t)-96\pi^2+(96\alpha k^2-6\alpha)\pi-48\beta k^2) \cosh^4(t) \\ \nonumber 
 &+& (-64k(\beta k^2+6\pi^2+\frac{3}{4}\pi\alpha)\sinh(t)-192\pi\alpha k^2+48\beta k^2+48\pi^2) \cosh^2(t) \\ \nonumber 
 &+& (64\beta k^3+192\pi^2 k)\sinh(t)+96\pi\alpha k^2]
\end{eqnarray}
\begin{equation} \label{omega}
\omega(t)=\frac{f(t)}{g(t)}
\end{equation}
Where the functions $f(t)$ and $g(t)$ are given as 
\begin{eqnarray} 
f(t)&=& (48\pi^2+6\pi\alpha-\beta)\cosh^6(t)\\ \nonumber
&+&((64(\pi^2+\frac{1}{4}\pi\alpha-\frac{1}{16}\beta))k\sinh(t)-160\pi^2+(-32\alpha k^2-22\alpha)\pi+16\beta k^2+4\beta)\cosh^4(t) \\ \nonumber
&+&((64(-2\pi^2-\frac{5}{4}\pi\alpha+\beta (k^2+\frac{1}{2})))k\sinh(t)+176\pi^2+(64\alpha k^2+16\alpha)\pi+48\beta k^2)\cosh^2(t) \\ \nonumber
&-&64k(\beta k^2-\pi^2-\pi\alpha)\sinh(t)-32\pi\alpha k^2-64\beta k^2-64\pi^2.
\end{eqnarray}
\begin{eqnarray} 
g(t)&=& (-48\pi^2-6\pi\alpha+\beta)\cosh^6(t) \\ \nonumber
&+&(-(192(\pi^2+\frac{1}{4}\pi\alpha-\frac{1}{16}\beta))k\sinh(t)+96\pi^2+(-96\alpha k^2+6\alpha)\pi+48\beta k^2)\cosh^4(t) \\ \nonumber
&+&(64k(\beta k^2+6\pi^2+\frac{3}{4}\pi\alpha)\sinh(t)+192\pi\alpha k^2-48\beta k^2-48\pi^2)\cosh^2(t) \\ \nonumber
&+&(-64\beta k^3-192\pi^2 k)\sinh(t)-96\pi\alpha k^2.
\end{eqnarray}
The jerk parameter in cosmology is defined as \cite{j1,j2} 
\begin{equation}\label{jerk}
j=\frac{\dddot{a}}{aH^3}=q+2q^2-\frac{\dot{q}}{H}
\end{equation}
where $q$ is the deceleration parameter and $\dddot{a}$ is the third derivative of the scale factor with respect to the cosmic time.
The jerk parameter provides a convenient method to describe models close to $\Lambda CDM$. The deceleration-to-acceleration cosmic transition happens for models with positive value of $j$ and negative value of $q$ \cite{paplo1}, flat $\Lambda CDM$ models have $j = 1$ \cite{j3}. For the current model we get the jerk parameter as
\begin{equation}
j=1+\frac{2}{\cosh^2(t)}.
\end{equation}
The variation of $q(t)$, $p(t)$, $\rho(t)$ and $\omega(t)$ versus cosmic time $t$ is shown in Fig. 1. The deceleration parameter $q$ shows a change in sign from positive (decelerating phase) to negative (accelerating phase). It starts at $q=1$ describing a decelerating radiation-dominated universe in a good agreement with the complete cosmic history investigated in \cite{peri}, passes the matter dominated universe at $q=\frac{1}{2}$ and ends with an accelerating universe at $q=-1$ (de Sitter universe). We have tried several positive, negative and zero values for $\alpha$ and $\beta$ with $k=0$, $1$ and $-1$. Some of these values are summerized in table $1$. Fig. 1(b) shows a wrong behavior of the energy density $\rho(t)$ where it goes to $-\infty$ as $t \rightarrow 0$, this behavior happens for all positive values of $\alpha$ and $\beta$ which means that positive values of $\alpha$ and $\beta$ are not allowed. All other plots in Fig. 1 have been plotted with $\alpha=\beta=-0.01$. The energy density in Fig. 1(c) is a positive decreasing function that goes to $+\infty$ as $t \rightarrow 0$. The evolution of the pressure is shown in Fig. 1(d), it starst as a positive decreasing function during the early-time where the expansion was decelerating, and then becomes negative in the late-time accelerating universe. It is generally believed that a negative pressure is required to achieve a repulsive gravity producing the accelerated expansion in the FRW cosmology. So, this behavior of the pressure in the current model agrees with the standard cosmological model where the early Universe ($z \rightarrow \infty$) is decelerating and filled with positive pressure component, while in far future with dark energy domination the expansion is accelerated \cite{stt}. It has also been shown that the positive pressure with viscosity leads to a decelerated expansion in the framework of the causal Israel-Stewart formalism \cite{460}. Setting $\alpha=\beta=0$ gives exactly the same behavior of the pressure and energy density illustrated in Fig. 1(c),(d).\par
\begin{table}[H]\label{tap}
\centering
\tiny
    \begin{tabular}{ | p{0.3 cm} | p{1cm} | p{1cm} | p{1cm} | p{1cm} | p{1cm} | p{1cm} | p{1cm} | p{1cm} | p{1cm} | p{1cm} | p{1cm} | }
    \hline
     $\alpha$ & 1 & 0.5 & 0.2&0.002 &0 & 0.02&0 &-0.5 &-0.01 &0&-0.2\\ \hline
    $\beta$ & 1 & 0.5 &0.1 &0.001 & 0.01& 0&0 &0 & 0&-0.01&-0.1\\ \hline
    $\rho(t)$ & $\rightarrow -\infty$ as $t \rightarrow 0$ &  $\rightarrow -\infty$ as $t \rightarrow 0$ & $\rightarrow -\infty$ as $t \rightarrow 0$ & $\rightarrow -\infty$ as $t \rightarrow 0$ &  $\rightarrow -\infty$ as $t \rightarrow 0$& +ve and $\rightarrow \infty$ as $t \rightarrow 0$ & +ve and $\rightarrow \infty$ as $t \rightarrow 0$ &$\rightarrow -\infty$ as $t \rightarrow 0$ & $\rightarrow -\infty$ as $t \rightarrow 0$ &+ve and $\rightarrow \infty$ as $t \rightarrow 0$&+ve and $\rightarrow \infty$ as $t \rightarrow 0$ \\ \hline
    \end{tabular}
		\caption {The behavior of $\rho(t)$ for different values of $\alpha$ and $\beta$. This behavior is the same for $k=0$, $1$ and $-1$. }
		\end{table}
In order to understand the nature of dark energy, it is essential to detect the value and evolution of the EoS parameter $\omega = \frac{p}{\rho}$. The value of $\omega$ is $0$ for dust, $1/3$ for radiation and $-1$ for vacuum energy (cosmological constant). In some scalar field models we can have $\omega \leq -1$ for phantom field and $-1 \leq \omega \leq 1$ for quintessence field. $\omega$ can evolve across the cosmological constant boundary $\omega = -1$ for quintom field. The largest value of $\omega$ consistent with causality is $\omega=1$ for some exotic type of matter called stiff matter \cite{zeld} where the speed of sound is equal to the speed of light. \par
The evolution of the EoS parameter $\omega(t)$ is shown in Fig. 1(e). We can see that for $k=0, 1$ the behavior agrees with recent observations which show the consistency of the cosmological constant dark energy scenario (a perfect fluid with $\omega=-1$). We also notice there is no evolution across the cosmological constant boundary $\omega=-1$ (the phantom divide line), i.e. there is no crossing to the phantom era $\omega < -1$. The conditions to have a phantom universe in a specific model have been investigated in \cite{nogo} where a no-go theorem has been suggested. According to this theorem, the EoS parameter of a single perfect fluid or a single scalar field can not cross the phantom divide line $\omega=-1$.  Since the modified cosmological equations (\ref{cosm1}) and (\ref{cosm2}) have been derived considering the perfect fluid matter as source in the universe, the proved no-go theorem explains the non-existence of the phantom behaviour in the current model.  \par
Since we are interested in finding the correct values of $\alpha$ and $\beta$ required for a stable flat universe in which a deceleration-to-acceleration transition occurs, we have examined the evolution of $\omega(t)$ for several values of these two prefactors (table 2). As we can see from table 2, a violation of causality ($\omega >1$) is allowed for all values of $\alpha$ and $\beta$ except for $\alpha=0$ and $\beta=0$ (Fig. 2(a)) where the maximum allowed value is $\omega$ $\approx \frac{1}{3}$. So, according to the current model, the evolution of the EoS parameter suggests zero values of $\alpha$ and $\beta$. This agrees with the analysis presented in \cite{zer} where it has been shown that the zero value is the unique choice consistent with both the holographic principle \cite{topo} and statistical mechanics.
\begin{table}[h!]\label{tap2}
\centering
\tiny
    \begin{tabular}{ | p{0.3 cm} | p{2cm} | p{2cm} | p{2cm} | p{2cm} | p{2cm} | p{2cm} | }
    \hline
     $\alpha$ & 0.02&0 &-0.5 &-0.01 &0     &-0.2\\ \hline
    $\beta$   & 0   &0 &0    & 0    &-0.01 &-0.1\\ \hline
		$\omega(t)$ & $-1 \leq \omega(t) \leq 1.65$ & $-1 \leq \omega(t) \leq 0.33$ & $-1 \leq \omega(t) \leq 1.85$ &  $-1 \leq \omega(t) \leq 1.7$ &  $-1 \leq \omega(t) \leq 3$& $-1 \leq \omega(t) \leq 3$\\ \hhline{|=|=|=|=|=|=|=|}
		  $\alpha$ & -0.8 & -1.5 &-2 &-3 & -4 &-4.5\\ \hline
    $\beta$   & -0.5 & -1.5 & -2 & -3 & -4 &-4.5\\ \hline
		$\omega(t)$ & $-1 \leq \omega(t) \leq 3.15$ & $-1 \leq \omega(t) \lesssim 3.15$ & $-1 \leq \omega(t) \leq 3.25$  & $-1 \leq \omega(t) \leq 3.5$  & $-1 \leq \omega(t) \leq 4$ & $-1 \leq \omega(t) \leq 4.7$  \\ \hline
    \end{tabular}
		\caption {The range of $\omega(t)$ for different values of $\alpha$ and $\beta$ ($k=0$).}
		\end{table}
Fig. 2(d) shows that the jerk parameter decreases with cosmic time until it becomes constant $j=1$ at late-time. So, at late-time of the universe the current entropy-corrected model tends to a flat $\Lambda CDM$ model .
\section{Stability of the model}
The physical acceptability of the current entropy-corrected model can be checked through testing the classical linear energy conditions \cite{ec11,ec12}, the sound speed, and the new nonlinear energy conditions \cite{ec,FEC1, FEC2, detec}. It has been shown that
the classical linear energy conditions (namely, the null $\rho + p\geq 0$; weak $\rho \geq 0$, $\rho + p\geq 0$; strong $\rho + 3p\geq 0$ and dominant $\rho \geq \left|p\right|$ energy conditions) should be replaced by other nonlinear energy conditions
in the presence of semiclassical quantum effects \cite{ec, detec}. It has also been pointed out that these linear energy conditions can not be valid in completely general situations and then they are not fundamental physics \cite{ec2,parc}. The nonlinear energy conditions we consider in this work are (i) The flux energy condition (FEC): $\rho^2 \geq p_i^2$ \cite{FEC1, FEC2}, first presented in \cite{FEC1} when obtaining  entropy bounds for uncollapsed systems. (ii) The determinant energy condition (DETEC): $ \rho . \Pi p_i \geq 0$ \cite{detec}. (ii) The trace-of-square energy condition (TOSEC): $\rho^2 + \sum p_i^2 \geq 0$ \cite{detec}. \par
The strong energy condition (SEC) expresses the `highly restrictive' statement that gravity should always be attractive. However, it has been shown that even in the classical regime this condition fails when describing the universe in the current accelerated epoch and during inflation \cite{ec3,ec4,ec5}. For the current model, and because the negative pressure represents a repulsive gravity, we don't expect the SEC to be fulfilled in the late-time epoch dominated by negative pressure (dark energy dominated epoch). We can see that clearly in Fig. 1(h) where the SEC becomes invalid during the late-time accelerating DE dominated epoch for all values of $k$. The dominant energy condition (DEC) expresses the fact that energy density should be non-negative and should propagate in a causal way. Fig. 1(i) shows that this condition is always valid for $k=0,1$. The same happens for the WEC (Fig. 1(g)) which is valid all the time for $k=0,1$. The behavior of the nonlinear energy conditions has been plotted in Fig. 1(j),(k),(l). Fig. 1(j) shows that the flux energy condition FEC is always valid for $k=0,1$, while Fig. 1(k),(l) show that the determinant energy condition DETEC and the trace-of-square energy condition TOSEC are always valid for all $k$ values. \par
The adiabatic square sound speed $c_s^2=\frac{dp}{d\rho}$ should be positive and less than $1$. This is because causality implies that sound speed must be less than the speed of light and so the condition $0 \leq \frac{dp}{d\rho} \leq 1$ should be always satisfied ($c=G=1$ in relativistic units). For the current model, we get
\begin{equation}
c_s^2 = \frac{f_1(t)}{g_1(t)}
\end{equation}
Where
\begin{eqnarray}
f_1(t)&=& \left(32 (\pi^2+\frac{1}{4} \pi \alpha-\frac{1}{16} \beta) k \sinh(t)+(-32 \pi \alpha+16 \beta) k^2-16 \pi^2-4 \pi \alpha+\beta\right) \cosh^4(t) \\  \nonumber
&+& \left((\beta k^2-\frac{2}{3}\pi^2-\frac{11}{12}\pi\alpha+\frac{5}{12}\beta)96k\sinh(t)+(64\pi\alpha+128\beta)k^2+32\pi^2-12\pi\alpha+8\beta\right)\times \\  \nonumber
&&\cosh(t)^2 -96\left(\beta k^2-\frac{1}{3}\pi^2-\frac{5}{6}\pi \alpha-\frac{1}{4}\beta\right)k\sinh(t)+(-32\pi \alpha-144 \beta)k^2-16\pi (\pi-\alpha).
\end{eqnarray}
And
\begin{eqnarray}
g_1(t)&=& \left(-96 (\pi^2+\frac{1}{4} \pi \alpha-\frac{1}{16} \beta) k \sinh(t)+(-96 \pi \alpha+48 \beta) k^2-16 \pi^2-4 \pi \alpha+\beta\right) \times \\  \nonumber
&&\cosh^4(t) + \left((\beta k^2+2\pi^2-\frac{1}{4}\pi\alpha+\frac{1}{4}\beta)96k\sinh(t)+ 192\pi( \alpha k^2+\frac{1}{2}\pi+\frac{1}{16}\alpha)\right)\times \\  \nonumber
&& \cosh(t)^2 -96\left(\beta k^2+\pi^2-\frac{1}{2}\pi \alpha\right)k\sinh(t)+(-96\pi \alpha-48 \beta)k^2-48\pi^2.
\end{eqnarray}
Fig. 1(f) shows that this stability condition is satisfied only for the flat universe case. Hence, in the current entropy-corrected model, the flat universe is the most stable one in a good agreement with recent observations which favor a flat universe.
\section{The total density parameter}
\subsection{Entropy-corrected holographic dark energy}
For a universe dominated by the holographic dark energy and the pressureless matter, the total density parameter $\Omega$ can be written as
	\begin{equation}
	\Omega=\Omega_{m}+\Omega_{\Lambda}
	\end{equation}
	Where $\Omega_m$ is the density parameter for dark matter, and $\Omega_{\Lambda}$ is the density parameter for dark energy. The energy density of the so-called entropy-corrected holographic dark energy (ECHDE) has been proposed by Wei in the form \cite{echde}.
\begin{equation}\label{echde}
\rho_{\Lambda}=3c^2M^2_pL^{-2}+\gamma L^{-4} \ln(M^2_pL^2)+\eta L^{-4},
\end{equation}
In units where $M_p^2=8\pi G=1$, $M_p$ is the reduced Planck mass, $L$ is the characteristic length scale (the IR
cut-off) of the system and $c$ is a constant. Some observations suggested that the value of $c$ is close to unity for a flat
universe \cite{Lib}. If $\gamma=\eta=0$, equation (\ref{echde}) gives the well-known holographic dark energy (HDE) density: 
\begin{equation}\label{hde}
\rho_{\Lambda}=3c^2M^2_pL^{-2},
\end{equation}
when $L$ is very small, the last two terms in (\ref{echde}) become comparable to the first one which means that these corrections make sense only at the early-time evolution of the universe. As $L$ increases, the universe gets larger and the ECHDE (\ref{echde}) reduces to the ordinary HDE (\ref{hde}). Since inflation happens at a very early stage of cosmic evolution, the last two terms in (\ref{echde}) are important during inflation epoch. Then, during the radiation and matter-dominated epochs the universe is large enough for such terms to be ignored. 
The simplest choice for the characteristic length scale $L$ is the Hubble scale $L=\frac{1}{H}$, which leads to energy density comparable to
the present-day dark energy \cite{le1, le2}. It has been suggested in \cite{hsu} that this choice cannot be used at late-times application since it cannot lead
to an accelerating universe. However, it has been shown in \cite{antihsu} that in a flat universe, and as soon as an interaction between dark energy and dark matter is taken into account, the choice $L=\frac{1}{H}$ can simultaneously drive accelerated expansion and solve the coincidence problem. An important feature of the current model is that the late-time cosmic acceleration is generated by ad-hoc mechanism (we generate it by hand through an empirical ansatz). Consequently, there are no worries about not obtaining the late-time cosmic acceleration in case of using the Hubble scale as the IR cutoff. It has been shown in \cite{exx} that, Contrary to the case of late-time application, the Hubble scale can be used in inflationary applications. Despite the drawbacks with the Hubble scale as the IR cutoff, it has been used in \cite{ir} to explore the properties of holographic ellipsoidal cosmologies and the possibility to develop observationally testable cosmologies has been proved. Since the flat universe is the most stable case in the current entropy-corrected model, we can now consider only a flat universe. Hence, $\Omega_{M}$ and $\Omega_{\Lambda}$ can be expressed as
\begin{equation}
\Omega_{m}=\frac{\rho(t)}{3H^2}=\frac{(48\pi^2+6\pi\alpha-\beta)\cosh^6t-6\pi(16\pi+\alpha)\cosh^4t+48\pi^2\cosh^2t}{384 \pi^3 \sinh^4(t)\cosh^2(t)} 
\end{equation}
\begin{eqnarray}
\Omega_{\Lambda}&=&\frac{3c^2H^2+\gamma H^4 \ln \frac{1}{H^2}+\eta H^4}{3H^2}\\  \nonumber
&=&\frac{1}{12\sinh^2(t) } \left[ \gamma \cosh^2(t)\ln\left(\frac{\sinh(t)}{\cosh(t)}\right)^2+(2 \gamma \ln(2)+\eta+12c^2)\cosh^2(t)-12c^2\right].
\end{eqnarray}
Some observations suggest that for a flat universe $c=0.818^{+0.113}_{-0.097}$ \cite{Lib}. In Fig. 2(b), we have plotted the evolution of the total density parameter $\Omega$ versus cosmic time $t$ for $c=0.818+0.113$, $c=0.818-0.097$ and $c=1$. we observe that $\Omega\rightarrow 1$ as $c \rightarrow 1$. For $c=1$, $\Omega=1$ at late-time in a good agreement with observations \cite{sp}. The ratio of dark
matter energy density to the dark energy density defines the coincidence parameter $r=\frac{\rho_{m}}{\rho_{\Lambda}}$:
\begin{equation}
r=\frac{ \left( 48\,{\pi}^{2}+6\,\pi\,\alpha-\beta \right)  \cosh^4 t- \left( 96\,{\pi}^{2}+6\,\pi\,\alpha
 \right)  \cosh^2t +48\,{\pi}^{2}
}{ 32\,\sinh^2t  \left( a \cosh^2t \,\ln  (\tanh^2t) + \left( 2\,  a \ln  2  +12\,{c}^{2}+b \right)  \cosh^2t   -12\,{c}^{2} \right) {\pi}^{3}
}
\end{equation}
While observations show that this ratio is constant at the present time \cite{coin}, the standard $\Lambda CDM$ model doesn't agree with this observations. Fig. 2(e) shows that the coincidence parameter varies at the early-time evolution and remains constant at the late-time evolution.
\subsection{Modified holographic Ricci dark energy}
The energy density of the so-called modified holographic Ricci dark energy density (MHRDE) has been suggested in \cite{de2}. In this dark energy proposal, the energy density is a function of the Hubble parameter $H$ and its time derivative. It is given by 
\begin{equation}
\rho_{\Lambda}=3(\alpha_1H^2+\beta_1\dot{H}+\gamma_1\ddot{H}H^{-1})
\end{equation}
Where the behavior of $\rho_{\Lambda}$ depends heavily on the parameters $\alpha_1$, $\beta_1$ and $\gamma_1$. This model reduces to the modified Ricci dark energy model \cite{de3, de4} for $\alpha_1=0$. For the current model, $\rho_{\Lambda}$ can be expressed as
\begin{equation}\label{MHRDE}
\rho_{\Lambda}=\frac{\alpha_1 \cosh^2(t)-2\beta_1+8\gamma_1}{\cosh^2(t)}
\end{equation}
The evolution of the total density parameter $\Omega=\Omega_M+\Omega_{MHRDE}$ has been plotted in Fig. 2(c) for different values of the parameters $\alpha_1$, $\beta_1$ and $\gamma_1$. We have found that $\Omega=1$ for $\alpha_1=\beta_1=\gamma_1=1$. The coincidence parameter r is given by:
\begin{equation} \label{coin}
r=-\frac{ \cosh^2t \left(  \left( {\pi}^{2}+
\frac{1}{8}\,\pi\,\alpha-\frac{\beta}{48} \right) \cosh^4t -2\,\pi\, \left( \pi+\frac{\alpha}{16} \right) \cosh^2t +{\pi}^{2} \right) 
}{16  \left( -\frac{1}{2}\,\alpha_{1}\, \cosh^2t 
+\beta_{1}-4\,\gamma_{1} \right) {\pi}^{3} \sinh^4t
}
\end{equation}
The behavior of (\ref{coin}) is shown in Fig. 2(f) .
\section{Conclusion}
A stable flat entropy-corrected FRW cosmological model has been constructed. The modified FRW cosmological equations have been solved using an ansatz $a(t)=A\sqrt{\sinh(\xi t)}$ which allows a deceleration-to-acceleration transition. We have found that this ansatz perfectly describes a stable flat universe for certain values of the parameters $\alpha$ and $\beta$. While positive values of $\alpha$ and $\beta$ don't lead to physically acceptable solutions, we can get physically acceptable solutions with negative and zero values.\par
The pressure in this model is positive during the early-time decelerating expansion, and negative during the late-time accelerating epoch dominated by dark energy. The evolution of $\omega(t)$ has been investigated for several values of $\alpha$ and $\beta$. We found that a violation of causality ($\omega >1$) is allowed for all values of $\alpha$ and $\beta$ except for $\alpha=\beta=0$ where $-1 \leq \omega(t) \lesssim \frac{1}{3}$.\par
Based on the stability analysis, the most stable solution is the flat one where only the flat case ($k=0$) passes the sound speed test plus the old classical and new nonlinear energy conditions. Another support for the flat universe in the current model comes from the jerk parameter where it tends to a flat $\Lambda$CDM with $j=1$ at late-time.\par
We have calculated the total density parameter $\Omega$ using two different proposals for dark energy. For the case of the entropy-corrected holographic dark energy, we found that $\Omega \rightarrow 1$ at late-time of cosmic evolution in a good agreement with observations. In case of considering the modified holographic Ricci dark energy, we found that $\Omega \rightarrow 1$ at late-time for $\alpha_1=\beta_1=\gamma_1=1$.\par
Finally, Some important previous results in literature must be considered. Entropy corrections can be related to the Shannon-Von Neumann entropy already discussed in cosmology \cite{note1, note2, note3}. In \cite{note1} it has been shown that the Von Neumann entropy can be compared to the thermodynamical entropy in FRW universe, it has also been found in \cite{note3} that the evaluation of the Shannon-Von Neumann entropy can be directly related to the cosmological constant. On the other hand, the adopted energy conditions can be related to a general approach recently proposed for modified theories of gravity \cite{note4,note5}. It has been shown that in spite of the energy conditions can be rewritten as in general relativity, their meaning can be totally different since the
causal structure, geodesic structure and gravitational interaction may be changed. Also, the energy conditions have to be considered in a wider sense where the validity of the inequalities does not guarantee the attractive nature of gravity. The modified gravity model proposed here can work for other higher-order models as reported in \cite{note6,note7}.

\begin{figure}[H]
  \centering            
  \subfigure[$q$]{\label{F63}\includegraphics[width=0.24\textwidth]{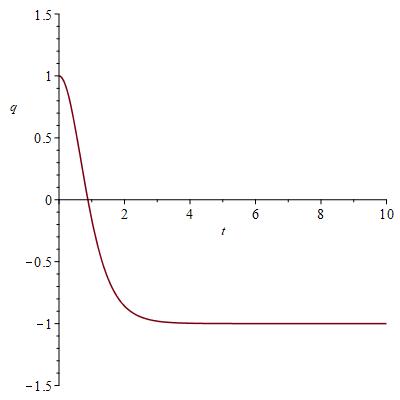}} 
	\subfigure[$\rho$ ( $\alpha$, $\beta$ $>0$).]{\label{F635}\includegraphics[width=0.24\textwidth]{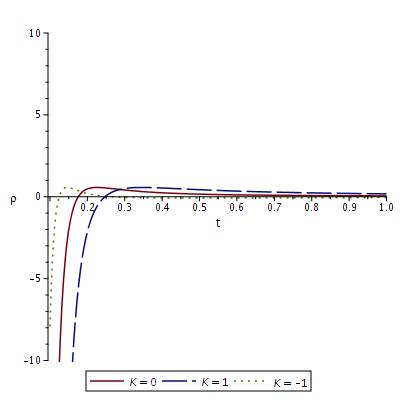}} 
	 \subfigure[$\rho$ ( $\alpha$= $\beta$ $=-0.01$).]{\label{F64}\includegraphics[width=0.24\textwidth]{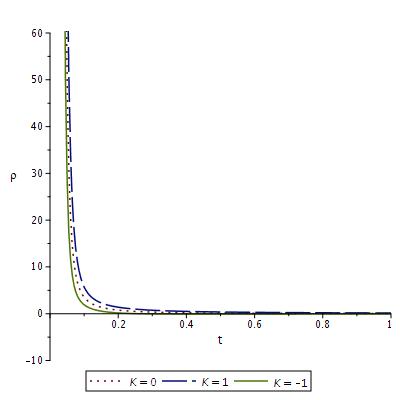}} 
	\subfigure[$p$]{\label{F423}\includegraphics[width=0.24\textwidth]{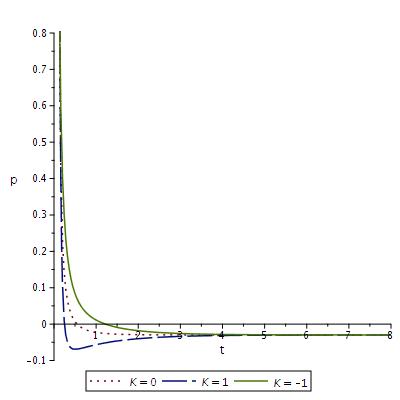}} \\
	\subfigure[$\omega$]{\label{F4}\includegraphics[width=0.24\textwidth]{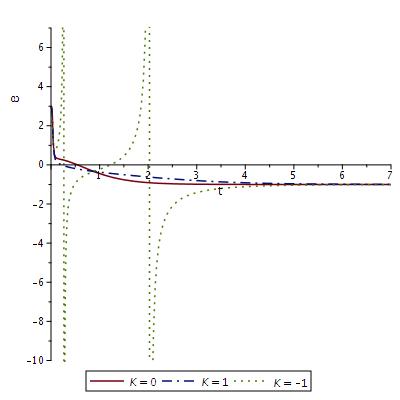}}
	\subfigure[$C_s^2$]{\label{F5}\includegraphics[width=0.24\textwidth]{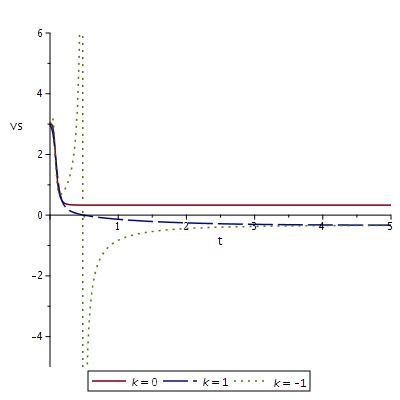}}
	\subfigure[$\rho+p$]{\label{F59}\includegraphics[width=0.24\textwidth]{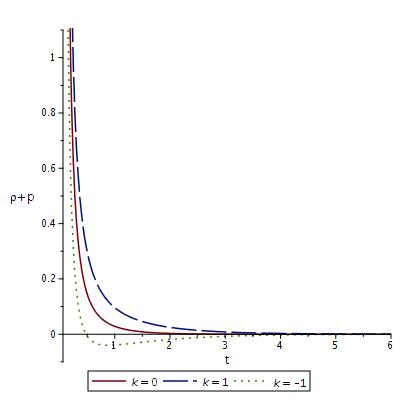}}
	\subfigure[$\rho+3p$]{\label{F597}\includegraphics[width=0.24\textwidth]{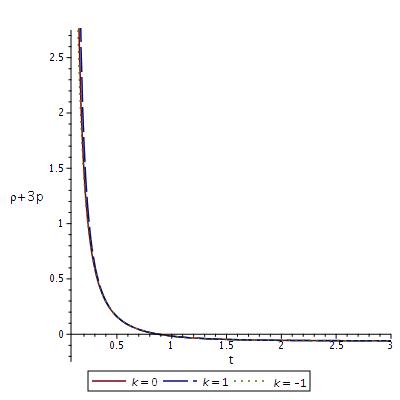}}\\
	\subfigure[$\rho-p$]{\label{F598}\includegraphics[width=0.24\textwidth]{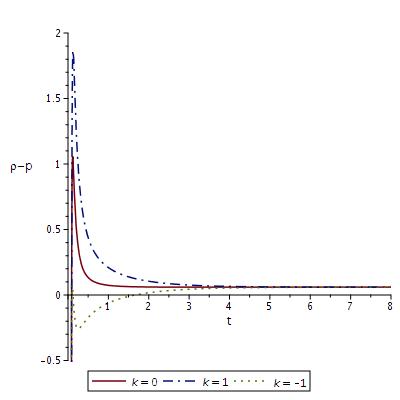}}
	  \subfigure[$\rho^2-p^2$]{\label{F636}\includegraphics[width=0.24\textwidth]{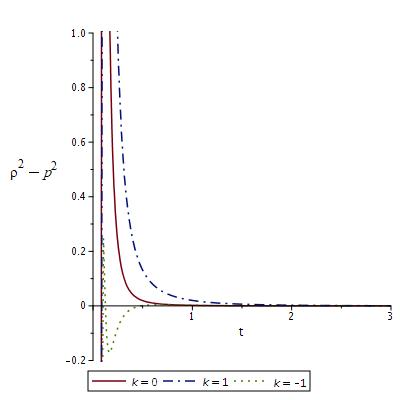}} 
	 \subfigure[$\rho . p^3$]{\label{F649}\includegraphics[width=0.24\textwidth]{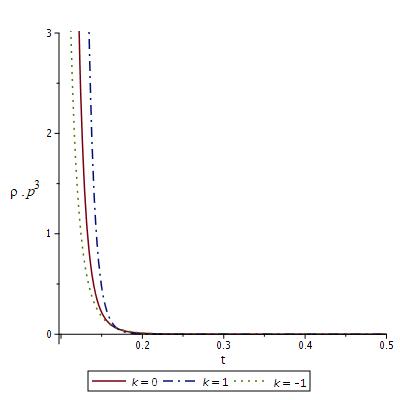}} 
	\subfigure[$\rho^2 . 3p^2$]{\label{F6498}\includegraphics[width=0.24\textwidth]{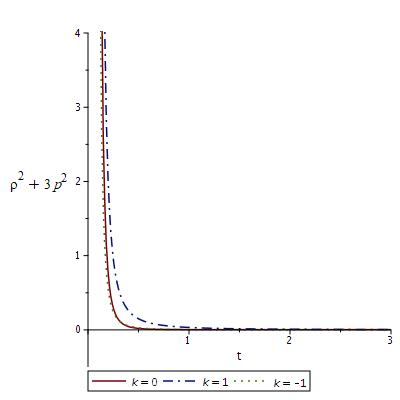}} 
  \caption{Fig. 1(a) The deceleration parameter varies in the range $-1\leq q \leq 1$. (b) A wrong behavior of the energy density $\rho(t)$ near the beginning of time for positive values of $\alpha$ and $\beta$, here $\alpha=\beta=1$. (c), (d), (e),(f), (g), (h) and (i) show the behavior of $\rho(t)$, $p(t)$, $\omega(t)$, $C_s^2$ and classical energy conditions for $\alpha=\beta=-0.01$. The same behavior in (c), (d), (f), (g),(h) and (i) has been obtained for $\alpha=\beta=0$. (j), (k) and (l) show the behaviour of nonlinear energy conditions for $\alpha=\beta=-0.01$ ( the same behavior have been obtained for $\alpha=\beta=0$).}
  \label{fig:cassimir55}
\end{figure}
\begin{figure}[H]
  \centering            
	\subfigure[$\omega$ ($\alpha=\beta=0$)]{\label{F488}\includegraphics[width=0.24\textwidth]{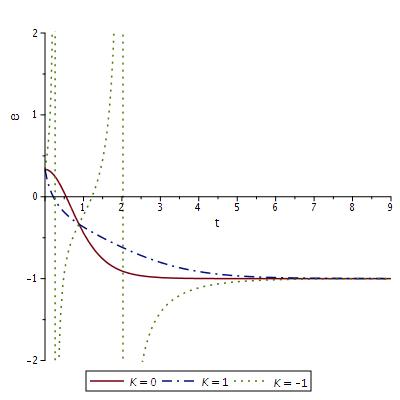}}
	 \subfigure[$\Omega=\Omega_M+\Omega_{ECHDE}$]{\label{F6396}\includegraphics[width=0.24\textwidth]{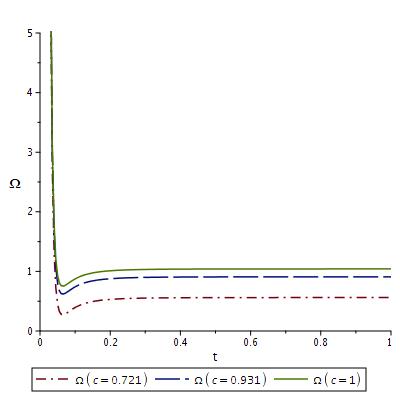}} 
	 \subfigure[$\Omega=\Omega_M+\Omega_{MHRDE}$]{\label{F606}\includegraphics[width=0.24\textwidth]{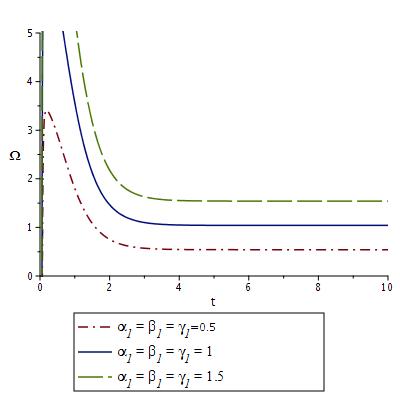}}
	 \subfigure[$j$]{\label{F6006}\includegraphics[width=0.24\textwidth]{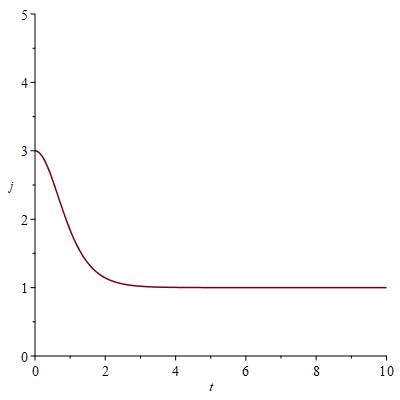}}\\
		 \subfigure[$$]{\label{F600886}\includegraphics[width=0.24\textwidth]{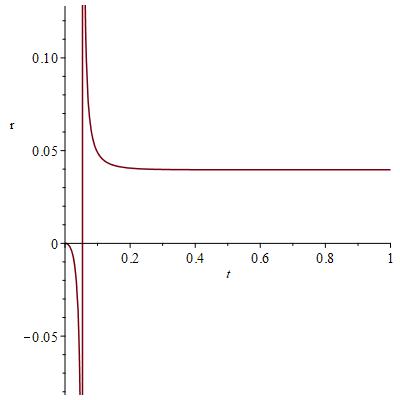}}
			 \subfigure[$$]{\label{F6000006}\includegraphics[width=0.24\textwidth]{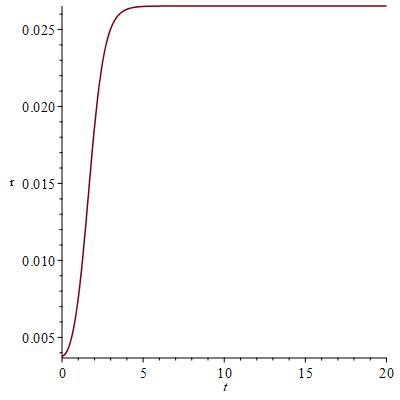}}
  \caption{ Fig. 2(a) shows that the EoS parameter varies in the range $-1 \leq \omega(t) \leq 0.33$ for $\alpha=\beta=0$. The variation of the total density parameter $\Omega$ considering the ECHDE is shown in (b) with $\gamma=\eta=0.01$, while The variation of $\Omega$ considering the MHRDE is shown in (c) for different values of $\alpha_1$, $\beta_1$ and $\gamma_1$. The evolution of the jerk parameter (d) shows that at late-time $j=1$ and the current stable flat entropy-corrected model tends to a flat $\Lambda CDM$ model. The behavior of the coincidence parameter for the ECHDE and MHRDE is shown in (e) and (f) respectively.}
  \label{fig:cassimir595}
\end{figure}

\end{document}